# Superpower E-Beam Sources and Performance Estimates for Compact THz FELs

Richard B. True, *Life Fellow, IEEE,* Henry P. Freund, Michael V. Fazio, *Life Member, IEEE,*
and Patrick G. O'Shea, *Fellow, IEEE*

*Abstract*—High peak and average power free-electron lasers (FELs) in the terahertz region (THz) require small diameter, low-emittance, and high voltage electron beams. This paper presents two 1.5-2 MV, 100-200 A, thermionic cathode electron source approaches for compact megawatt range peak power, multi-kilowatt average power, high repetition rate, THz FELs. The preferred beam generation system includes grading electrodes and is quite compact compared to the other standard diode gun approaches. Both provide highly compressed beams at the waist having low values of normalized rms emittance. In particular, the new injector approach with grading electrodes, operating at a body voltage of 1.5-2 MV and 100 A, has a normalized rms emittance of roughly 10 μm at the beam waist. Power supply switching considerations are considered in the paper as well as considerations for very high-voltage multi-stage depressed collectors for device efficiency enhancement. Based on these designs, we provide performance estimates for FELs operating in the THz spectral region.

## I. INTRODUCTION

IN a recent paper by the authors, design issues associated with developing compact free-electron lasers (or free-electron masers) capable of producing high peak and average power radiation in the THz region were addressed [1]. This paper builds upon the principles set forth in that paper by presenting a more detailed look at the electron source design. Depressed collector and power supply issues are also discussed. Based upon these electron gun designs, we also update the performance estimates discussed in that earlier work.

Typically, workers in the field assume the THz region extends from 300 GHz to 10 THz (some take 100 GHz as the lower frequency end). In this paper, we focus on a 1 THz device recognizing that the basic design can be scaled to operate at higher or lower frequency to satisfy certain potentially useful terrestrial, high-altitude, research, or space applications. These may include systems that take advantage of the atmospheric windows, powerful high-resolution imaging systems, or potential other systems for nuclear, biological, and chemical applications.

In general, FELs do not need very high field strength magnets like in gyro-devices but do require MeV energy electron beams to reach THz frequencies. Various technologies exist for producing MeV electron beams, however, the thermionic megavolt electron beam sources presented in this paper appears uniquely suited for THz FELs producing peak powers in the megawatt range and 10 kilowatts of average power or more.

## II. ELECTRON INJECTOR BASICS

Many FELs at the top end of the THz band or higher utilize photo-cathode-driven electron guns and RF accelerators. To reach, for instance, wavelengths of 1-3 μm or shorter, energies above 40 MeV, and normalized rms beam emittances less than about 7 μm, are required [2]. This combination invariably results in relatively short 1-10 ps electron bunches from the injection system.

It is possible to build an FEL at the lower end of the THz band using a RF photo-injector recognizing that there is no need for the full RF linac since 1-4 MeV beam energies are achievable from the RF gun. However, this approach is typically large and complex requiring a drive laser and high-power RF source such as a klystron with all its associated power supplies. In addition, the short electron bunches produced by this technology mean that slippage in the wiggler can limit performance.

Since high-performance low band THz FELs can employ beams with normalized rms emittances of 10 – 20 μm, the ultra-low emittances of RF photo-injectors are not required. It is thus possible to utilize thermionic cathode guns and beam formation systems for high repetition rate, high peak and average power FELs. High-quality low-emittance thermionic cathode sources are quite robust and are ideally suited for rugged compact FELs. Further, since the beam from the thermionic emitter is continuous during the pulse, the interaction is similar to that in the classic Ubitron [3] which is known to have relatively high DC to RF conversion efficiency.

The objective beam parameters for the 1 THz FEL of this paper are given in [1]. First, the beam voltage range is 1.5 to 2

Richard B. True was formerly with L-3 Communications, Electron Devices Division, San Carlos, CA 94070. He is currently an independent scientific advisor (email: true86b@gmail.com).

Henry.P. Freund is with the University of Maryland, College Park, MD 20742, the University of New Mexico, Albuquerque, New Mexico, 87131, and NOVA Physical Science and Simulations, Vienna, VA 22182 (email: freundh0523@gmail.com)

Michael V. Fazio is with the SLAC National Accelerator Laboratory, Menlo Park, CA 94025 (email: mfazio@slac.stanford.edu )

Patrick G. O'Shea is with the University of Maryland, College Park, MD 20742 (email: poshea@umd.edu )

MV and the current range is nominally 100 A to 200 A. The injected electron beam then enters the assumed two-plane focusing wiggler near the beam focal point location. In the case of a 1.5 MV, 100 A, 10 mm-mrad normalized rms emittance beam, for capture and matching, the beam diameter should lie between 0.826 and 1.168 mm for a magnetic field magnitude of $1.0 - 2.0$ kG, and a wiggler period of $1.0 - 2.0$ cm [1]. The beam in this case is comfortably smaller than the gap needed to generate these wiggler fields for excellent propagation of the beam through the wiggler. For instance, in Fig. 5 of [1], wiggler fields between $1.0 - 2.0$ kG can be generated using gaps between $7 - 9$ mm. In this case, ratio of the average matched beam diameter (1 mm) to the average gap (8 mm) is approximately 1/8.

Finally, the electron sources in this paper are all thermionic dispenser cathodes which are known to be very reliable and provide long life [4]. Cathode loading in the guns herein is commensurate with or below that in the standard SLAC 5045 klystron gun. Note that cathode loading in the 5045 under normal operating conditions is 6.5 A/cm$^2$ (beam voltage 350 kV, beam current 413 A, 9.0 cm diameter cathode). The pulse length in the klystron is 3.5 µs and the pulse repetition frequency is 180 Hz.

## III. ELECTRON GUN APPROACHES

The beam from the standard non-convergent diode gun design of Fig. 6 in [1] is virtually perfect. That is, cathode current density versus radius and beam current density versus radius downstream are uniform, and the beam has virtually no spherical aberration. At 2 MV, the surface gradient of the focus electrode is conservative for reliable high voltage standoff during the pulse.

This gun with a slightly simplified anode is shown in Fig. 1 and which we refer to as the "Big20" gun. It provides essentially identical results to the original. With no magnetic lens field applied, the emittance remains essentially constant through the anode opening and downstream. This is shown in Fig. 3 where we plot the 95% beam envelope and the normalized rms emittance versus the axial distance from the cathode for the guns shown in Figs. 1 and 2. Note that all the gun simulations herein were performed using the fully-relativistic and well-validated code DEMEOS [5] where the finite-emittance thermal beam in the code is represented by a 3-ray model using definitions of emittance presented in [6] – [8].

Figure 2 corresponds to the gun of Fig. 10 in [1] with zero focusing lens field and which we refer to as the "GradEl" gun. It includes a mod anode for current switching and control, and three grading electrodes to achieve higher voltage standoff to the body in a minimum amount of space [9]. Since beam current varies with anode voltage in this case according to the 3/2-power law, $I_0 = 10^{-6} P_\mu V_a^{3/2}$, and since the anode perveance for this gun is 2 µperv, raising the anode voltage to 195 V results in a beam current equal to 173 A which is close to the gun in Fig. 1. Conclusions for the 153 A case pertain to this case as well.

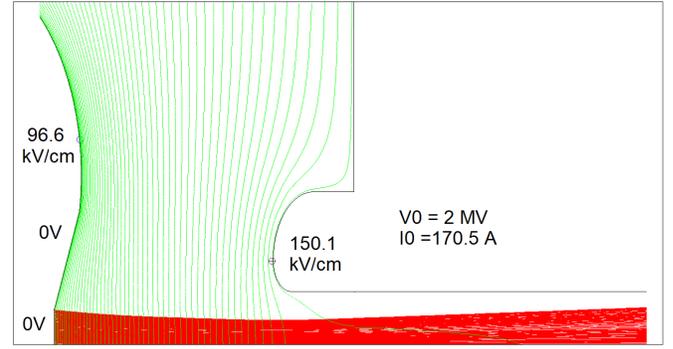

**Fig. 1.** Computer simulation of 2 MV, 170.5 A, 0.060 µperv "Big20" diode gun with no magnetic focusing lens (external region not shown and radial scale expanded in all gun plots).

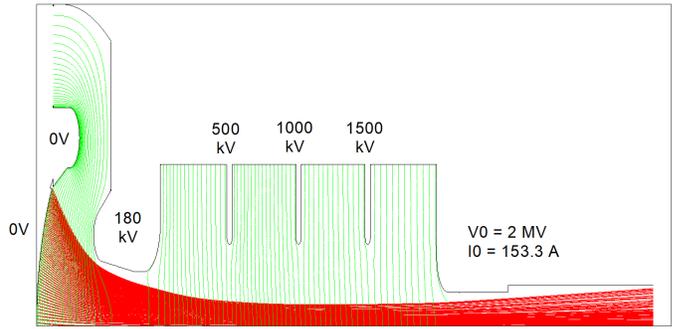

**Fig. 2.** Computer simulation of 2 MV, 153.3 A, 0.055 µperv Pierce gun plus acceleration section with grading electrodes ("GradEl" gun) with no magnetic focusing lens.

The first observation in the plot of Fig. 3 is that the slope of the spreading 95% beam envelope beyond the anode opening is similar for the Big20 and GradEl guns despite the fact that the beam in Fig. 2 is much smaller than that in Fig. 1. The beam size and slope both impact the focal length after the magnetic lens.

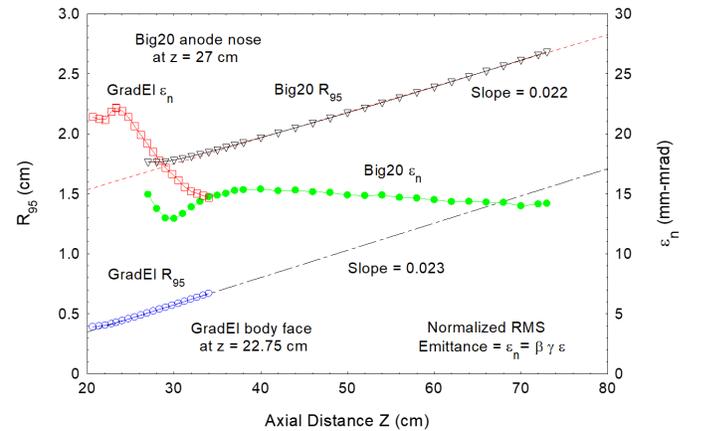

**Fig. 3.** Plot of the radii enclosing 95% of the beam current and normalized rms emittance versus distance from the cathode center for the guns of Figs. 1 and 2 (cathode disk radius = 2.54 cm in both cases).

TPS*

The second observation is that for both cases, the normalized rms emittance, $\varepsilon_n$, for both guns stabilizes downstream to a value below 15 μm. Note that the unnormalized rms emittance, $\varepsilon$, is computed from Eq. A-3 in [7] in DEMEOS, then the normalized rms emittance from $\varepsilon_n = \beta\gamma\varepsilon$ where $\beta = \upsilon/c$ and $\gamma$ is the relativistic mass factor. For example, for the right most point of the 2 MV Big20 gun case in Fig. 3, $\varepsilon = 2.96$ μm and $\varepsilon_n = 14.2$ μm.

## IV. DIODE GUN PLUS FOCUSSING LENS

While the normalized emittances of the gun designs in the preceding section are considered quite good from the standpoint of FEL performance at THz frequencies, we now consider the effect on emittance and other parameters when the beams are magnetically compressed prior to injection into the wiggler.

The electron gun of Fig. 1 with a standard magnetic focusing coil situated downstream of the anode body face was analyzed first. The general configuration of this type of "thin" lens is shown in Fig. 6.3 (and Fig. 1.6 for the offset gap case) in [10] noting the variation of the magnetic field along the axis can be assumed to vary as $B_z = B_0 \text{sech}^2 (1.32z/a_{pp})$ where $a_{pp}$ is the inner radius of the polepieces [11].

While this type of lens can focus the the beam down to the appropriate size using this type of lens, spherical aberration causes emittance to increase markedly downstream over the magnetic field free case as illustrated in Fig. 7 where we plot the 95% beam envelope and the normalized rms emittance versus the axial distance from the cathode for the gun described in this section. Note that we tried various combinations of lens spacing, strength, and pole piece radius without success.

Fortunately, we were able to achieve success with another type of lens, a "thick" magnetic lens similar to a short solenoid. Since the beam is so perfect, the lens must be essentially perfect too with no spherical aberration to preserve the low value of emittance of the unfocused beam.

The computer simulation of the beam from the diode gun focussed by such a solenoidal lens is shown in Fig. 4. Clearly, in this case, as shown in Fig. 7, the normalized rms emittance remains low throughout the beam compression phase and is similar to that of the beam in the field free case.

A generic picture of this well-known lens type is shown in Fig. 7.1 of [10] or Fig. 5 – 7 of [12] for example. In such lenses, it is reasonable to assume the axial input and output magnetic field varies according to [13]. Solenoid design and analysis is discussed in more detail in, for example, [5] pp. 548-551.

A summary of various results for this gun and focusing arrangement is shown in Table I. In general, the beam is adequate for a 1 THz FEL in all the cases shown. Note that lower normalized rms emittance results from reducing the voltage from 2 MV to 1.5 MV and lowering the lens field to preserve the focusing. If the gun is physically scaled by 0.8, at 1.5 MV and 112 amps, the focussed beam is a little smaller and the normalized rms emittance under 11 mm-mrad at the waist.

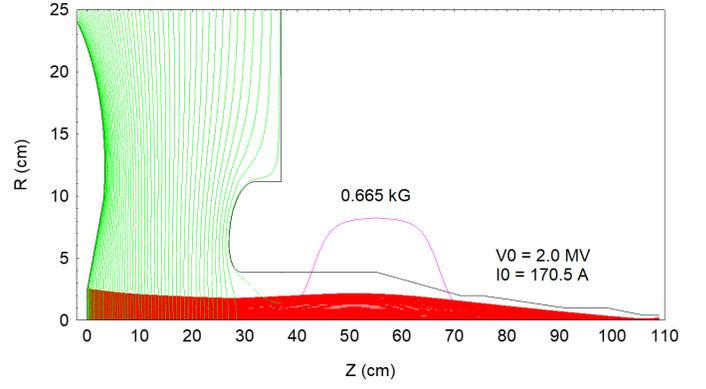

**Fig. 4.** Computer simulation of 2 MV, 170.5 A, 0.060 μperv Big20 diode gun with 0.665 kG peak solenoidal magnetic focusing "thick" lens.

Table I.
Big20 Focussed Beam Summary (Thick Lens)

| Parameters | Z | Rrms[1] | R95 | $\varepsilon_n$ | eV Spd[2] |
|---|---|---|---|---|---|
| | (cm) | (cm) | (cm) | mm-mrad | (%) |
| V0 = 2.0 MV | 107.0 | 0.051 | 0.077 | 15.3 | 0.09 |
| Rc = 2.54 cm | 108.0 | 0.035 | 0.061 | 15.0 | 0.17 |
| I0 = 170.5 A | 109.0 | 0.030 | 0.051 | 16.9 | 0.81 |
| Bpk = 0.665 kG | | | | | |
| V0 = 1.5 MV | 107.0 | 0.044 | 0.077 | 14.6 | 0.11 |
| Rc = 2.54 cm | 108.0 | 0.035 | 0.065 | 14.3 | 0.13 |
| I0 = 112.1 A | 109.0 | 0.041 | 0.077 | 14.2 | 0.18 |
| Bpk = 0.526 kG | | | | | |
| V0 = 1.5 MV | 85.6 | 0.035 | 0.061 | 11.2 | 0.11 |
| Rc = 2.03 cm | 86.4 | 0.027 | 0.051 | 11.0 | 0.14 |
| I0 = 112.1 A | 87.2 | 0.032 | 0.059 | 11.4 | 0.21 |
| Bpk = 0.670 kG | | | | | |

[1]Rrms = sqrt <R^2>   [2]eV spread = std. dev. eV / <eV>

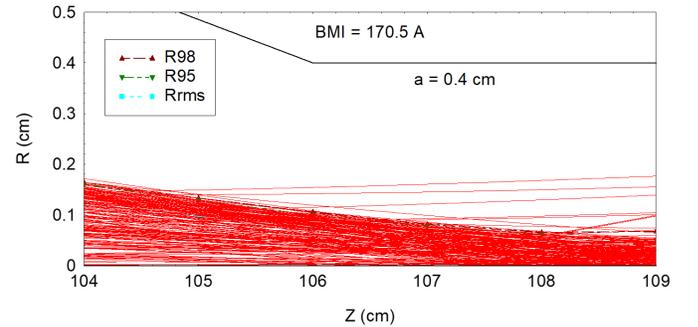

**Fig. 5.** Zoom view of focussed beam near the waist for the case of Fig. 4. Appearance of focussed beam in other cases qualitatively similar to this.

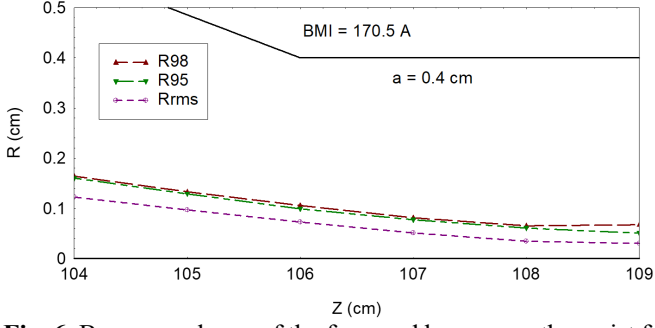

**Fig. 6.** Beam envelopes of the focussed beam near the waist for the case of Fig. 4.

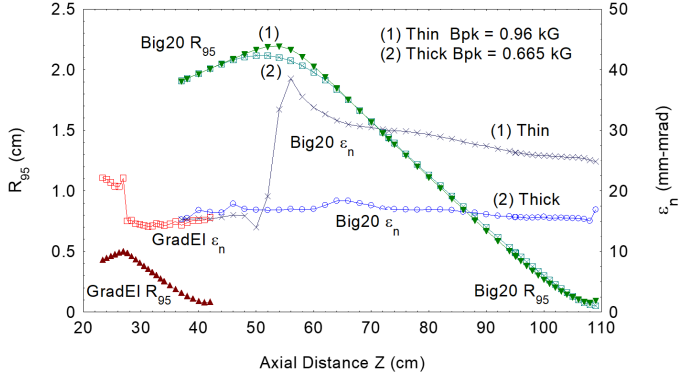

**Fig. 7.** Comparison of the 95% radius and normalized rms emittance versus distance for the Big20 gun with thin (1) and thick (2) focusing lenses and for the GradEl gun (cathode disk radius = 2.54 cm and V0 = 2 MV for cases shown).

## V. PIERCE GUN PLUS ACCELERATION SECTION WITH FOCUSSING LENS

The diode gun of Section IV provides an excellent beam for use in compact THz FELs. On the other hand, it can be seen in Fig. 7 that the distance to the beam waist in the gun of this section is significantly less. The precompression of the beam from the Pierce gun of Fig. 2 enables the shorter focal length but it also helps in the achievement of excellent normalized rms emittance at the waist. This type of injector is thus ideally suited for use in *very* compact THz FELs. There are many other advantages in this basic approach which will become apparent in the following discussions and quantitative data.

The baseline compressed beam gun which includes grading electrodes is shown in Fig. 8 (which in Ref. [1] is similar to Fig. 10 scaled to have a cathode diameter the same as that in Fig. 8 of [1] or 2.032 cm). It is based upon a scaled version of the SLAC 5045 electron gun followed by an acceleration section which includes grading electrodes to break up the gap voltage.

In this case, the divergent electrostatic lens through the body orifice causes spherical aberration where the outer portion of the beam is deflected increasingly more strongly than the inner portion. In the design, this is compensated for by the spherical aberration from the "thin" magnetic lens which tends to be relatively stronger away from the axis. This compensation system tends to recover the low value of emittance as illustrated by the unscaled baseline case in Fig. 7.

Table II shows the quantitative results for Fig. 8. The rms beam radius at the waist at z = 32.8 cm is 0.034 cm, the radius enclosing 95% of the beam current, $R_{95}$, is 0.049 cm, and the normalized rms emittance, $\varepsilon_n$, is 9.7 μm.

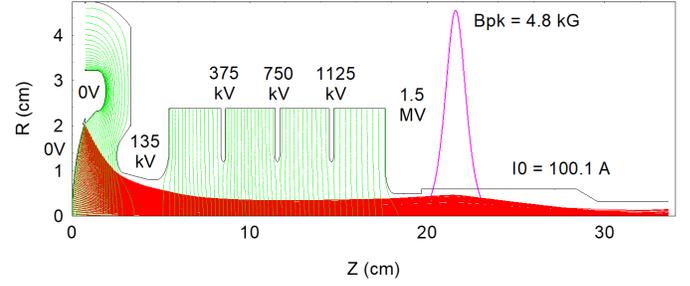

**Fig. 8.** Computer simulation of a 0.8 scale 1.5 MV, 100.1 A, 0.055 μperv grading electrode gun with 4.8 kG peak magnetic focusing lens.

Table II.
GradEl Focussed Beam Summary (Thin Lens)

| Parameters | Z (cm) | Rrms (cm) | R95 (cm) | $\varepsilon_n$ mm-mrad | eV Spd (%) |
|---|---|---|---|---|---|
| V0 = 2.0 MV | 40.0 | 0.057 | 0.086 | 15.1 | 0.16 |
| Rc = 2.54 cm | 41.0 | 0.050 | 0.076 | 15.2 | 0.14 |
| Va = 180 kV | 42.0 | 0.048 | 0.081 | 15.6 | 0.37 |
| I0 = 153.3 A | | | | | |
| G1 = 500 kV | | | | | |
| Bpk = 4.8 kG | | | | | |
| V0 = 2.0 MV | 38.0 | 0.069 | 0.075 | 11.1 | 0.24 |
| Rc = 2.54 cm | 39.0 | 0.052 | 0.058 | 11.1 | 0.22 |
| Va = 135 kV | 40.0 | 0.044 | 0.071 | 10.9 | 0.24 |
| I0 = 100.1 A | | | | | |
| G1 = 320 kV | | | | | |
| Bpk = 4.8 kG | | | | | |
| V0 = 1.5 MV | 40.0 | 0.056 | 0.087 | 12.1 | 0.14 |
| Rc = 2.54 cm | 41.0 | 0.047 | 0.070 | 12.0 | 0.13 |
| Va = 135 kV | 42.0 | 0.044 | 0.065 | 11.9 | 0.13 |
| I0 = 100.1 A | | | | | |
| G1 = 375 kV | | | | | |
| Bpk = 3.8 kG | | | | | |
| V0 = 1.5 MV | 32.0 | 0.039 | 0.061 | 9.8 | 0.14 |
| Rc = 2.03 cm | 32.8 | 0.034 | 0.049 | 9.7 | 0.14 |
| Va = 135 kV | 33.6 | 0.035 | 0.055 | 9.6 | 0.16 |
| I0 = 100.1 A | | | | | |
| G1 = 375 kV | | | | | |
| Bpk = 4.8 kG | | | | | |

To explore field-free beam flow beyond the waist, a mirror image of the region to the left of the original problem-closing Neumann boundary at z = 33.6 cm was set up, and the simulation was continued downstream. The rms radius of the beam in this case is essentially symmetric about this plane as illustrated in Fig. 9.

It can also be seen in Fig. 9 that beam divergence beyond the waist is low especially in view of the huge aspect ratio of the plot. Further, the normalized rms emittance remains low beyond the waist very far downstream helping to make beam capture and control by the wiggler magnetic field much easier.

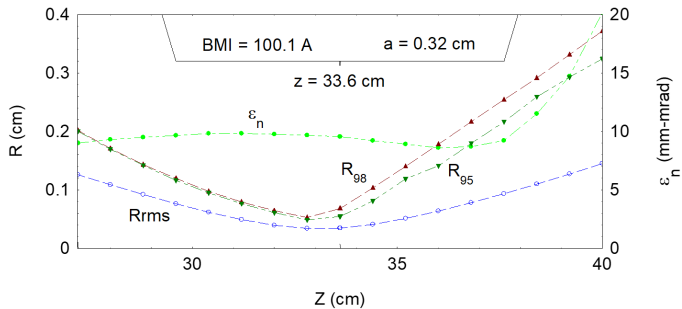

**Fig. 9.** Zoom view of focussed beam near the waist and beyond for the 0.8 scale 1.5 MV GradEl gun.

## VI. COMMENTS ON RESULTS

Both gun and beam focusing approaches provide beams that can be used in 1 THz FELs. The diode gun approach is more conventional and when paired with a thick magnetic focusing lens, an appropriately sized low-emittance beam is available for injection into the wiggler. The Pierce gun with grading electrodes is more novel and also provides a beam small enough to conform to the requirements of a 1 THz FEL with appropriately low normalized rms emittance.

As can be seen in Fig. 8, the compressed beam gun with grading electrodes represents a much more compact configuration. This has the further advantage of easier beam alignment. Since the 95% beam compression is over 1500 for both approaches, beam alignment is quite important.

The peak magnetic field level in the focusing lens of the GradEl gun is substantially higher than in the thick lens of the Big20. Fabrication of the thick lens solenoid should be relatively straightforward which is an advantage. On the other hand, the diode gun is more sensitive to the level of magnetic field in comparison to the compressed beam gun with grading electrodes.

In both designs, beam spread with distance is very low making the beams easy to capture in two-plane focusing wigglers designed for operation at 1 THz. The beam filling factor into the wiggler is very small in each case so beam interception on the circuit should be minimal over the full interaction region. It would be useful to explore both beam capture and transport through the wiggler using a fully three-dimensional code such as MICHELLE [14] or BOA [15] for a full exploration and validation of the predicted results.

It is possible to adjust the size and location of the beam waist in both injectors by simply adjusting the magnetic focusing lens field strength. This represents an important handle from an experimental point of view. Further, in the GradEl gun, the voltages of the electrode stages can be adjusted to fine-tune overall injector performance as shown in Table II and Fig. 10.

In the diode gun, when the focusing magnetic field is fixed, the focus position shifts from its normal position near the entrance of the wiggler to a position upstream of the wiggler as the voltage is lowered. The beam spreads out before the wiggler which can create a heat load during turn-on and turn-off. As long as the transient phase is short enough and there is adequate body cooling, this may represent an advantage in protecting the circuit.

Quantitatively, for instance, for the 2 MV case with a lens field of 0.665 kG, reducing the voltage to 1.5 MV lowers the emitted current to 112 A whereupon the beam focus shifts to 84 cm. This point is shown in Fig. 10. Note the 95% beam size is 0.045 cm and the normalized rms emittance is 18.6 mm-mrad. At $z = 106$ cm, the assumed entrance to the wiggler, because of the overfocussed spreading of the beam, only 5.9% of the total beam current makes it into the wiggler itself.

As above, this may not be a problem if the transient phase is short enough. Achievement of rise and fall times having a rough order of magnitude of 100 ns from a high repetition rate modulator should be safe but does represents a challenge. It may be possible to sync the lens field to keep the focal spot fixed under these conditions but this also represents a challenge. Another potential choice is to lower the magnetic field so the focal point is situated deep within the wiggler so when it moves back during the transient phase it will still lie within an acceptable distance from the wiggler entrance. This requires further detailed study.

On the other hand, in the gun with grading electrodes, voltages can be applied before application of the mod anode voltage used to switch the beam on and off. Given the relatively low modulating anode voltage level, very fast switching is possible in this case.

Fig. 10 shows that the location of the focal spot remains reasonable for capture by the wiggler (the entrance of which is assumed to be at $z = 37$ cm) over a wide range of current for a fixed lens field and constant body voltage.

If the voltage of one or more of the grading electrodes is trimmed in concert with the mod anode voltage, shown in Figs. 10, 11 and Table II, it is possible to provide a beam which can be focused from low current up to full current. This is an important feature insofar as it can reduce the thermal heat load by the entrance of the wiggler and the production of unwanted x-rays.

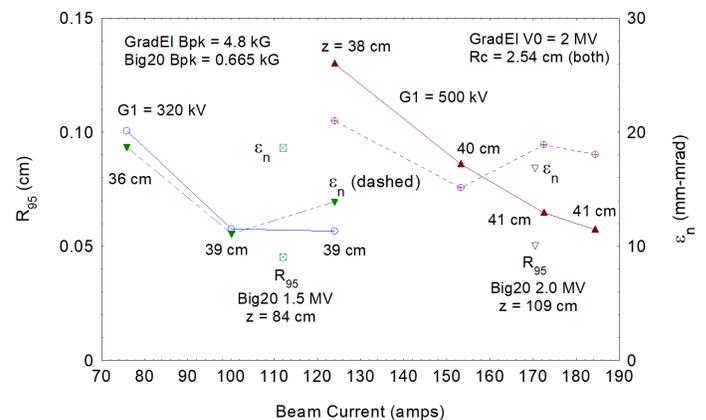

**Fig. 10.** 95% beam radius and normalized rms emittance over range of current for the Big20 and GradEl guns.

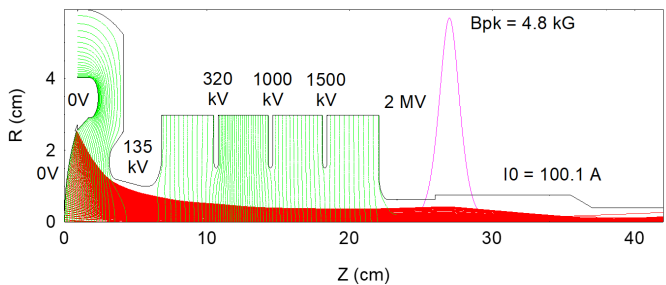

**Fig. 11.** Electrode voltages can be adjusted to provide excellent performance over a wide range of current in the grading electrode injector.

## VII. MECHANCAL ISSUES

The diode gun in Fig. 1 can be mounted within a large high-voltage vacuum enclosure comprised of a stack of ceramic disks and grading electrodes to break up the voltage and shield the insulator rings themselves. It would be preferable to mount the gun coaxially within the cylindrical vacuum enclosure in the standard "flashlight" configuration used in most klystrons [4], [5]. However, in this case, it might be possible for the insulator stack to end at the anode plate in order to save space. Outside of the vacuum assembly there needs to be enough distance to the wall of the oil tank or pressure vessel if that is used to stand off the full level of pulse voltage. The insulator assembly may be domed or tapered to help accomplish this.

One example of a very high voltage gun mounting structure housing a 3.2 MV diode gun is the one used in the dual-axis radiographic hydrodynamic test facility (DARHT-II) accelerator injector at Los Alamos [16] - [19]. This system is overly large for a compact FEL but includes many important concepts for reliably standing off pulse voltages in the megavolt range. In any event, the overall size of the enclosure for the diode gun herein will be substantial for reliable 2 MV operation.

On the other hand, it should be possible to mount the GradEl gun of Fig. 2 in a novel compact enclosure as shown in Fig. 12. In this case, the high voltage is broken into four separate high voltage regions. The only point where the full voltage is seen in the design is along the axis where the gradients can be arranged to be conservative for operation at 2 MV. The outer wall of the enclosure in this case resembles an oil barrel.

The figure shows flat plates but in an actual practical gun, we envision the flat plates to transition to low angle cones a distance away from the axis to prevent "oil canning." Also, the plate-to-plate and plate-to-anode spacings can be increased by such a move for even more reliable high voltage standoff.

The basic support structure in Fig. 12 can be reduced in size for operation at 1.5 MV using the scaled gun. On the other hand, we think that it may be best to use the 2 MV design since it provides a greater margin of safety against breakdown and more experimental flexibility.

Recall in earlier sections, for the 100 A, 2.54 cm radius cathode cases, the normalized rms emittance for this injector is just over 10 mm-mrad at both 2 and 1.5 MV. Also, the larger spacings of the 2 MV design should allow the pulse length to be over 3 times longer at 1.5 MV based on theory [20] using $k$ (a function of pulse length) in the equation $V = k L^{0.8}$ expressed as a series in a spreadsheet used for voltage standoff calculations [21].

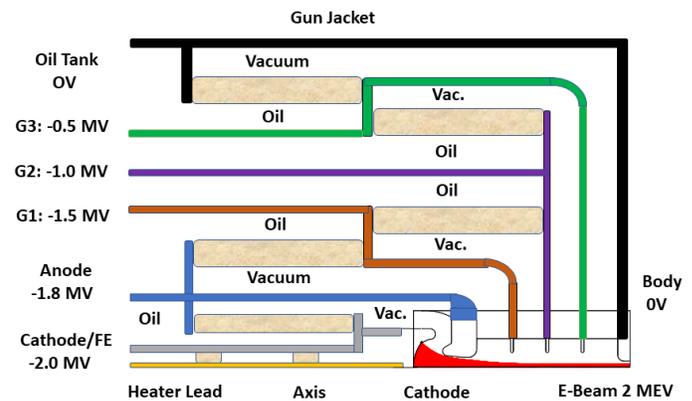

**Fig. 12.** Simplified sketch of Pierce gun plus three grading electrodes in compact support structure (from [1] Fig. 10). High voltage region is subdivided into four isolated lower voltage regions in this design. In the figure, insulators are shown in tan, corona rings by end of the insulators and other such details are not shown. The plot is unscaled.

Further, the use of a longer pulse enables a lower repetition rate for a given level of average output power resulting in a lower level of transient power lost to the body during the pulse turn-on and turn-off phase.

## VIII. SPENT BEAM COLLECTION

A further aspect of a compact FEL is the inclusion of a multistage depressed collector to recover most of the spent beam energy. What we envision is a set of collector plates sandwiched between ceramic disk insulators mounted within a large oil tank in which the oil represents the cooling fluid. The insulating oil is also used to provide high voltage standoff to the tank.

The collector plates inside, which are progressively more negative downstream so as to decelerate the beam, would be shaped for maximum collection efficiency and arranged to partially cover the ceramic insulators as is known in the state of the art. The final most negative element at the end can be a Faraday cage collector held at a potential say 5 percentage points positive with respect to the negative potential of the cathode.

A simplified sketch of a TWT two-stage depressed collector with rear Faraday cage element is shown in [5] p. 559 and [22] contains a useful chapter on collectors. Arranging the four high-voltage decks and anode modulator to drive both the gun and collector elements should be possible using modern fast solid state switching elements.

Alternately, it might be possible to use the arrangement of Fig. 12 turned around for a minimum volume collector. The front of the Faraday cage collector might take the place of the back of the anode in the figure where the cathode elements would be all dropped and the bucket would close at the end.

The energy distribution and spread of the spent beam from the wiggler impacts the openings and spacing of the plates, and the opening and configuration of the Faraday cage collector. They all will have to be carefully designed from an operational and power handling point of view. In this case, the basic overall concept may result in another key element in the development of highly compact high-power THz FELs.

## IX. FEL DESIGN CONSIDERATIONS

As in our previous work [1], we consider an FEL/FEM configured as a single-pass source based upon self-amplified spontaneous emission (SASE) due to the dearth of high-power seeds at THz frequencies. However, a high-gain/low-$Q$ oscillator configuration is a viable alternative. The first requirement in such a study is to determine the beam energy required to achieve resonance at the desired wavelength, which we choose to be 300 μm. The optimal choice for the electron gun is the GradEl gun described above (see Table II) with a beam energy of 2.0 MeV. Using a plane-polarized wiggler, the resonance at 300 μm can be achieved with a period of 1.5 – 2.0 cm and an on-axis amplitude of 0.5 – 2.0 kG. As shown in Table 2, using the GradEl gun we can achieve an rms energy spread of 0.2% and normalized emittances of 9.6 – 11.1 μm for beam voltages between 1.5 – 2.0 MV. This is close to the assumed emittance of 10 μm used previously [1] and the performance calculations using an emittance of 11 μm for a 2.0 MeV electron beam are similar to what was found previously.

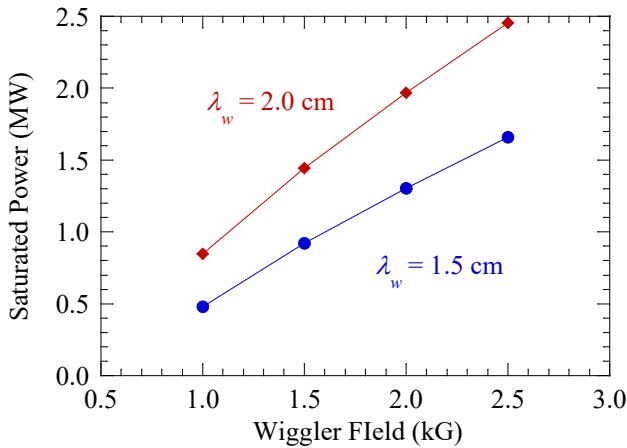

**Fig. 13.** Variation in the saturated power versus wiggler field strength for periods of 1.5 and 2.0 cm.

The results of calculations showing the gain length and saturated power [23] for a range of on-axis wiggler field strengths of 1.0 – 2.5 kG and periods of 1.5 and 2.0 cm are shown in Fig. 13. Observe that the saturated power ranges from about 0.5 – 2.5 MW.

The spontaneous noise power associated with the electron beam in this configuration is close to 56 μW; hence, the wiggler lengths associated with the cases shown in Fig. 13 range from 2 – 5 m which is reasonably compact.

As mentioned previously, the GradEl gun can produce flat-top pulses of between 1 – 2 μs at a repetition rate of 1 kHz corresponding to a duty factor of between 0.001 – 0.002. Given the pulse powers shown in Fig. 13, this implies that the average power ranges from 500 W – 5.0 kW.

## X. SUMMARY AND CONCLUSION

In this paper we have presented two promising thermionic cathode electron beam injector approaches that appear useful for compact 1 THz FELs. Combined with new solid-state pulsed power technology, we envision peak powers in the MW range, average powers of 5 kW or more, at repetition rates that can be as high as 1 kHz or more, from such devices.

In a series of detailed computer simulations and analyses, we have shown that high current and voltage, highly compressed, low emittance beams, suitable for use in THz FELs, can be created. Specifically, we have achieved 1.5-2 MV, 100 A and higher beams that are appropriately sized for injection into 1 THz wigglers. For all the 100 A cases, the 95% area compression ratios are greater than 1500, and in the case of the gun with grading electrodes of Fig. 2, normalized rms emittance values are just over 10 μm at the focused beam waist.

While this paper focused on compact thermionic cathode high-power injectors for FELs operating at 1 THz, it is possible to scale the components described herein to devices that operate at lower frequencies. In many respects, FELs designed to operate at lower frequency or lower powers become easier to achieve. Further, both of the injectors described herein may be useful in other accelerator or vacuum electron device applications whereas they can be readily scaled in size and power to satisfy the requirements of many other such devices.

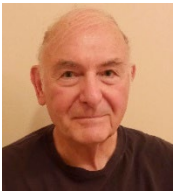

**Richard B. True** (Life Fellow, IEEE) received the Sc.B. in EE from Brown University in 1966, the M.S. in Microwave Engineering, and Ph.D. in Electrophysics, from the University of Connecticut in 1968 and 1972, respectively. He was the Chief Scientist at the Electron Devices Division of Litton, subsequently Northrop Grumman, and then L-3 Communications, for 25 years until his retirement in 2016. Over the years, he designed most of the electron beam optical systems in the travelling wave tubes, klystrons, and other devices manufactured by EDD using theories and software that he originated including DEMEOS. He has published numerous papers, he holds many patents, and he has received numerous IEEE and other awards. Three awards of special note are an IEEE Third Millennium Medal, the John R. Pierce Award for Excellence in Vacuum Electronics, and a first-year winner of the L-3 Communications Corporate Best Engineer Award. Dr. True served as Chair of the IEEE Vacuum Electronics Technical Committee from 2013-2016 and remains a member of this committee.

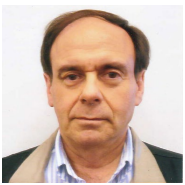

**Henry P. Freund** is a theoretical plasma physicist currently involved in studies of coherent radiation sources such as free-electron lasers (FELs), traveling wave tubes, klystrons and inductive output tubes, and cyclotron, Cerenkov, and orbitron masers by both analytical and numerical methods. He has published more than 180 papers in refereed journals, made numerous contributions to books and published proceedings, and coauthored a book entitled *Principles of Free-electron Lasers* [Springer, Cham, Switzerland, 2018, 3rd edition]. In addition to this scholarly activity, Dr. Freund has also made contributions to more popular scientific literature with contributions on free-electron lasers published in *Scientific American* magazine and the Academic Press Encyclopedia of Science and Technology. The article in *Scientific American* has been translated and republished in *Veda a Technika* (Science and Technology), which was published by the Czechoslovak Academy of Sciences. Dr. Freund has extensive experience in the simulation of microwave tubes. He has developed important simulation codes for treating free-electron lasers and masers, cyclotron masers and gyrotrons, traveling wave tubes, and klystrons and inductive output tubes.

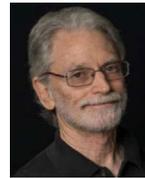

**Michael V. Fazio** (Life Member, IEEE) received his BSEE (1974), MEE (1975), and Ph.D. (1979) in electrical engineering from Rice University, Houston, TX, USA. He is currently Staff Emeritus at SLAC National Accelerator Laboratory. From 2015-2020 he served as the SLAC Associate Laboratory Director leading the Technology Innovation Directorate comprising RF accelerator research, quantum science & technology, advanced instrumentation, and detection systems development from microwave through gamma-ray wavelengths. From 1978-2010 at the Los Alamos National Laboratory, he held research and leadership positions including Division Director of the Intelligence, Space & Response Division responsible for satellite-based nuclear explosion detection and treaty monitoring instrumentation, planetary exploration and space science, proliferation detection, astrophysics, remote sensing, information science, and directed energy. He also served as Program Director for Space Situational Awareness and the National Security & Civilian Space Programs. Prior to 2005 he was a co-founder and Group Leader for the Los Alamos High Power Electrodynamics Group conducting high power RF source, free electron laser, advanced accelerator, and compact pulsed power R&D. His background in high power electrodynamics encompasses experimental research and development on advanced accelerators, high power RF sources, free-electron lasers, and compact pulsed power. In 2019 he chaired the DOE sponsored *Basic Research Needs Workshop on Compact Accelerators for Security and Medicine*. Dr. Fazio is the recipient of numerous awards including the US Energy Secretary's Achievement Award "in recognition of the accomplishments of the National Virtual Biotechnology Laboratory Team for mobilizing the research capabilities of the DOE National Laboratories to meet the challenges posed by the COVID-19 pandemic", the *Award for Excellence for Counter-Proliferation*, and the Achievement Award from the NASA Administrator as a Mars Curiosity ChemCam Instrument Development and Science Team member.

TPS*

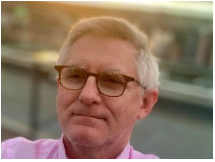
**Patrick G. O'Shea** (Fellow, IEEE) is professor of Electrical and Computer Engineering and Principal Investigator in the Bright Beams Collective Research Group at the University of Maryland. He was born in Cork, Ireland, and holds a BSc degree in Experimental Physics from the National University of Ireland, and an M.S. and Ph.D. in Physics from the University of Maryland. Professor O'Shea's technical expertise lies in the field of applied electromagnetics, nonlinear dynamics, and charged particle beam technology, and applications. He is a Fellow of the following societies: Royal Society for the Encouragement of Arts, Manufactures and Commerce; American Association for the Advancement of Science, American Physical Society; Institute of Electrical and Electronic Engineers; Irish Academy of Engineering. Professor O'Shea has previously served as: President of University College Cork, Ireland; Vice President and Chief Research Officer, Chair of the Department of Electrical & Computer Engineering, and Director of the Institute for Research in Electronics and Applied Physics at the University of Maryland; Faculty member at Duke University, and a Project Leader at the University of California Los Alamos National Laboratory.